**Correlation-preserving mean plausible values as a basis for prediction**

**in the context of Bayesian structural equation modeling**

André Beauducel* & Norbert Hilger

*University of Bonn, Germany*

## Abstract

Mean plausible values can be computed when Bayesian structural equation modeling (BSEM) is performed. As mean plausible values do not preserve the inter-factor correlations, they yield path coefficients that are different from the estimated path coefficients of the model. As it might be of interest to perform exactly the same prediction on the level of plausible values that has been estimated by BSEM, correlation-preserving mean plausible values were proposed. An example for the computation of the correlation preserving mean plausible values is given and the corresponding syntax is given in the Appendix.

* Corresponding author, adress for correspondence:

University of Bonn, Institute of Psychology, Kaiser-Karl-Ring 9, 53111 Bonn

Phone: +49228 734151, Email: beauducel@uni-bonn.de

Bayesian structural equation modeling (BSEM) has been proposed by Muthén and Asparouhov (2012) as an alternative to conventional structural equation modeling and several improvements of BSEM have meanwhile been proposed and realized (Asparouhov & Muthén, 2021; Asparouhov, Muthén, & Morin, 2015; Zitzmann & Hecht, 2019). Advantages of BSEM are that it can be performed with relatively small samples (Bonafede, Chiorri, Azzolina, 2021) and that priors for the variability of loadings can be specified. BSEM thereby allows to overcome problems of specifying fixed zero loadings in the independent clusters model (Beauducel & Hilger, 2020) and it also allows for the specification of complex loading patterns as, for example, circumplex models (Weide, Scheuble, & Beauducel, 2021).

As BSEM becomes more and more popular, the interest for score estimates of the latent variables or factors in these models may also increase. Asparouhov and Muthén (2010a, b) proposed mean plausible values as factor score estimates in the context of BSEM. Luo and Dimitrov (2018) found that even less than 500 imputations may be used in order to get mean plausible values with an appropriate validity. Moreover, Beauducel and Hilger (in press) have shown that mean plausible values of the exogenous factors ($\mathbf{P}_\xi$) based on 500 imputations have nearly the same coefficient of determinacy as the best linear factor score estimate, which is also termed regression factor score ($\mathbf{F}_\xi^R$). As $\mathbf{F}_\xi^R$ has the maximum determinacy, the result of Beauducel and Hilger (in press) implies that $\mathbf{P}_\xi$ based on more than 500 imputations is a proxy of $\mathbf{F}_\xi^R$.

Using $\mathbf{P}_\xi$ in the context of BSEM could be especially interesting in the context of the prediction of endogenous factors by exogenous factors. In applied settings, individuals might be selected according to their individual scores on exogenous factors (predictors). The selection of individuals according to their scores requires that the scores are valid indicators of the latent predictors which implies that they represent the underlying prediction model quite well. However, Skrondal and Laake (2001) have shown that using $\mathbf{F}_\xi^R$ for exogeneous factors $\xi$ and $\mathbf{F}_\eta^R$ for endogenous factors $\eta$ yields path coefficients that do not correspond to the path coefficients estimated by means of structural equation modeling. As $\mathbf{P}_\xi$ and $\mathbf{P}_\eta$ based on a large number of imputations are proxies of $\mathbf{F}_\xi^R$ and $\mathbf{F}_\eta^R$, it is expected that $\mathbf{P}_\xi$ and $\mathbf{P}_\eta$ yield biased path coefficients and inter-factor correlations (i.e., that the coefficients do not correspond to the coefficients obtained by means of BSEM). It could, however, be of interest to compute factor score estimates that allow for exactly the same predictions as the corresponding BSEM. The aim of the present study was therefore to provide a method that allows to transform $\mathbf{P}_{\xi}$ into scores resulting in path coefficients corresponding exactly to the path coefficients estimated by means of BSEM. After some definitions we provide the transformations for the correlation preserving mean plausible values and for the computation of their determinacy. Then, we provide an example based on a simulated data set to show the difference between conventional and correlation-preserving mean plausible values. The syntax for the transformation is given in the Appendix.

**Definitions**

We use the notation of Skrondal and Laake (2001), Jöreskog (1977), and Jöreskog and Sörbom (1989) with the latent regression model

$$\boldsymbol{\eta} = \boldsymbol{\Gamma}\boldsymbol{\xi} + \boldsymbol{\zeta}, \tag{1}$$

where $\boldsymbol{\Gamma}$ is the matrix of path coefficients for the prediction of the endogenous factors $\boldsymbol{\eta}$ by the exogeneous factors $\boldsymbol{\xi}$ and $\boldsymbol{\zeta}$ are the residuals, with $E(\boldsymbol{\zeta}\boldsymbol{\xi}^{'}) = \mathbf{0}$. The measurement model for the exogenous factors is

$$\mathbf{x} = \boldsymbol{\Lambda}_{\mathbf{x}}\boldsymbol{\xi} + \boldsymbol{\delta}, \tag{2}$$

where $\mathbf{x}$ represents the observed variables of the exogenous factors, $\boldsymbol{\Lambda}_{x}$ is the matrix of factor loadings, and $\boldsymbol{\delta}$ are the unique factors, with $E(\boldsymbol{\delta}\boldsymbol{\delta}^{'}) = diag(E(\boldsymbol{\delta}\boldsymbol{\delta}^{'})) = \boldsymbol{\Theta}_{\boldsymbol{\delta}}$, $E(\boldsymbol{\delta}\boldsymbol{\xi}^{'}) = \mathbf{0}$, and $E(\boldsymbol{\xi}\boldsymbol{\xi}^{'}) = \boldsymbol{\Phi}$, with $diag(E(\boldsymbol{\Phi})) = \mathbf{I}$, so that

$$E(\mathbf{x}\mathbf{x}^{'}) = \boldsymbol{\Sigma}_{\mathbf{x}} = \boldsymbol{\Lambda}_{\mathbf{x}}\boldsymbol{\Phi}\boldsymbol{\Lambda}_{\mathbf{x}}^{'} + \boldsymbol{\Theta}_{\boldsymbol{\delta}}. \tag{3}$$

The measurement model for the endogenous factors is

$$\mathbf{y} = \boldsymbol{\Lambda}_{\mathbf{y}}\boldsymbol{\eta} + \boldsymbol{\varepsilon}, \tag{4}$$

where $\mathbf{y}$ represents the observed variables of the exogenous variable, $\boldsymbol{\Lambda}_{y}$ is the matrix of factor loadings, and $\boldsymbol{\varepsilon}$ are the unique factors, with $E(\boldsymbol{\varepsilon}\boldsymbol{\varepsilon}^{'}) = diag(E(\boldsymbol{\varepsilon}\boldsymbol{\varepsilon}^{'})) = \boldsymbol{\Theta}_{\boldsymbol{\varepsilon}}$, $E(\boldsymbol{\delta}\boldsymbol{\xi}^{'}) = \mathbf{0}$, and $E(\boldsymbol{\eta}\boldsymbol{\eta}^{'}) = \boldsymbol{\Gamma}\boldsymbol{\Phi}\boldsymbol{\Gamma}^{'} + \boldsymbol{\Psi}$, with $diag(E(\boldsymbol{\eta}\boldsymbol{\eta}^{'})) = \mathbf{I}$, so that

$$E(\mathbf{y}\mathbf{y}^{'}) = \boldsymbol{\Sigma}_{\mathbf{y}} = \boldsymbol{\Lambda}_{\mathbf{y}}(\boldsymbol{\Gamma}\boldsymbol{\Phi}\boldsymbol{\Gamma}^{'} + \boldsymbol{\Psi})\boldsymbol{\Lambda}_{\mathbf{y}}^{'} + \boldsymbol{\Theta}_{\boldsymbol{\varepsilon}}. \tag{5}$$

## Correlation-preserving mean plausible values

The combined matrix of all factors is $\mathbf{F} = \begin{bmatrix} \boldsymbol{\xi} \\ \boldsymbol{\eta} \end{bmatrix}$, so that

$$\mathbf{C} = E(\mathbf{F}\mathbf{F}') = E\left(\begin{bmatrix} \boldsymbol{\xi} \\ \boldsymbol{\eta} \end{bmatrix} \begin{bmatrix} \boldsymbol{\xi}' \vdots \boldsymbol{\eta}' \end{bmatrix}\right) = E\left(\begin{bmatrix} \boldsymbol{\xi}\boldsymbol{\xi}' & \boldsymbol{\xi}\boldsymbol{\eta}' \\ \boldsymbol{\eta}\boldsymbol{\xi}' & \boldsymbol{\eta}\boldsymbol{\eta}' \end{bmatrix}\right). \tag{6}$$

As all factors have unit variance, $\mathbf{C}$ is a correlation matrix. According to Equations 1 to 5 the elements of $\mathbf{C}$ can be computed from the model parameter estimates

$$\mathbf{C} = \begin{bmatrix} \boldsymbol{\Phi} & \boldsymbol{\Phi}\boldsymbol{\Gamma}' \\ \boldsymbol{\Gamma}\boldsymbol{\Phi} & \boldsymbol{\Gamma}\boldsymbol{\Phi}\boldsymbol{\Gamma}' + \boldsymbol{\Psi} \end{bmatrix}. \tag{7}$$

The regression factor score $\mathbf{F}_\xi^R$ does not preserve the correlations in $\boldsymbol{\Phi}$ which follows from the covariances $E(\mathbf{F}_\xi^R \mathbf{F}_\xi^{R'}) = \boldsymbol{\Phi}\boldsymbol{\Lambda}_x' \boldsymbol{\Sigma}_x^{-1} \boldsymbol{\Lambda}_x \boldsymbol{\Phi}$ (Skrondal & Laake, 2001, Eq. 9), so that the inter-correlation of the regression factor scores is

$$\mathbf{C}_{\mathbf{F}_\xi^R \mathbf{F}_\xi^{R'}} = diag(\boldsymbol{\Phi}\boldsymbol{\Lambda}_x' \boldsymbol{\Sigma}_x^{-1} \boldsymbol{\Lambda}_x \boldsymbol{\Phi})^{-1/2} \boldsymbol{\Phi}\boldsymbol{\Lambda}_x' \boldsymbol{\Sigma}_x^{-1} \boldsymbol{\Lambda}_x \boldsymbol{\Phi} \, diag(\boldsymbol{\Phi}\boldsymbol{\Lambda}_x' \boldsymbol{\Sigma}_x^{-1} \boldsymbol{\Lambda}_x \boldsymbol{\Phi})^{-1/2}. \tag{8}$$

Inserting Equation 8 instead of $\boldsymbol{\Phi}$ into Equation 7 results in biased estimates for $E(\boldsymbol{\eta}\boldsymbol{\eta}')$ and $E(\boldsymbol{\eta}\boldsymbol{\xi}')$. As $\mathbf{P}_\xi$ and $\mathbf{P}_\eta$ based on a large number of imputations are proxies of $\mathbf{F}_\xi^R$ and $\mathbf{F}_\eta^R$, it follows that intercorrelations and path coefficients that are based on $\mathbf{P}_\xi$ and $\mathbf{P}_\eta$ will be biased.

Let $\mathbf{P} = \begin{bmatrix} diag(\mathbf{P}_\eta \mathbf{P}_\eta')^{-1/2} \mathbf{P}_\eta \\ diag(\mathbf{P}_\xi \mathbf{P}_\xi')^{-1/2} \mathbf{P}_\xi \end{bmatrix}$ and $\mathbf{C}_\mathbf{P} = E(\mathbf{P}\mathbf{P}')$ so that mean plausible values preserving the correlations in $\mathbf{C}$ can be defined as

$$\mathbf{P}_\mathbf{C} = \mathbf{C}^{1/2} \mathbf{C}_\mathbf{P}^{-1/2} \mathbf{P}, \tag{9}$$

where "$^{1/2}$" denotes the symmetric square-root, $\mathbf{P}_\mathbf{C} = \begin{bmatrix} \mathbf{P}_{\mathbf{C}\boldsymbol{\eta}} \\ \mathbf{P}_{\mathbf{C}\boldsymbol{\xi}} \end{bmatrix}$ and

$E(\mathbf{P}_\mathbf{C} \mathbf{P}_\mathbf{C}') = E(\mathbf{C}^{1/2} \mathbf{C}_\mathbf{P}^{-1/2} \mathbf{P}\mathbf{P}' \mathbf{C}_\mathbf{P}^{-1/2} \mathbf{C}^{1/2}) = \mathbf{C}^{1/2} \mathbf{C}_\mathbf{P}^{-1/2} \mathbf{C}_\mathbf{P} \mathbf{C}_\mathbf{P}^{-1/2} \mathbf{C}^{1/2} = \mathbf{C}.$

If the mean pausible values approximate the regression factor score, Equation 9 can be computed directly from the model parameters. For convenience, this is only illustrated for the exogenous factors, although it also holds for the endogenous factors.

For exogenous factors Equation 9 can be written as

$$\mathbf{P}_{C\xi} = \mathbf{\Phi}^{1/2} \mathbf{C}_{\mathbf{P}\xi}^{-1/2} diag(\mathbf{P}_{\xi}\mathbf{P}_{\xi}')^{-1/2} \mathbf{P}_{\xi}. \tag{10}$$

For $\mathbf{P}_{\xi} = \mathbf{F}_{\xi}^{R}$ it is possible to insert the right hand side of $diag(\mathbf{F}_{\xi}^{R}\mathbf{F}_{\xi}^{R'})^{-1/2}\mathbf{F}_{\xi}^{R}$ $= diag(\mathbf{\Phi}\mathbf{\Lambda}_{x}'\mathbf{\Sigma}_{x}^{-1}\mathbf{\Lambda}_{x}\mathbf{\Phi})^{-1/2}\mathbf{\Phi}\mathbf{\Lambda}_{x}'\mathbf{\Sigma}_{x}^{-1}$ for $diag(\mathbf{P}_{\xi}\mathbf{P}_{\xi}')^{-1/2}\mathbf{P}_{\xi}$ and the right hand side of Equation 8 for $\mathbf{C}_{\mathbf{P}\xi}^{-1/2}$ in Equation 10. This yields

$$\mathbf{P}_{C\xi} =$$
$$\mathbf{\Phi}^{1/2}(diag(\mathbf{\Phi}\mathbf{\Lambda}_{x}'\mathbf{\Sigma}_{x}^{-1}\mathbf{\Lambda}_{x}\mathbf{\Phi})^{-1/2}\mathbf{\Phi}\mathbf{\Lambda}_{x}'\mathbf{\Sigma}_{x}^{-1}\mathbf{\Lambda}_{x}\mathbf{\Phi}\,diag(\mathbf{\Phi}\mathbf{\Lambda}_{x}'\mathbf{\Sigma}_{x}^{-1}\mathbf{\Lambda}_{x}\mathbf{\Phi})^{-1/2})^{-1/2}diag(\mathbf{\Phi}\mathbf{\Lambda}_{x}'\mathbf{\Sigma}_{x}^{-1}\mathbf{\Lambda}_{x}\mathbf{\Phi})^{-1/2}\mathbf{\Phi}\mathbf{\Lambda}_{x}'\mathbf{\Sigma}_{x}^{-1}\mathbf{x},$$
$$\tag{11}$$

so that no mean plausible values but model parameters and the measured variables are needed to compute the correlation preserving plausible values. For $\mathbf{\Phi} = \mathbf{I}$ Equation 11 can be transformed to

$$\mathbf{P}_{C\xi} = diag(\mathbf{\Lambda}_{x}'\mathbf{\Sigma}_{x}^{-1}\mathbf{\Lambda}_{x})^{1/4}(\mathbf{\Lambda}_{x}'\mathbf{\Sigma}_{x}^{-1}\mathbf{\Lambda}_{x})^{-1/2}diag(\mathbf{\Lambda}_{x}'\mathbf{\Sigma}_{x}^{-1}\mathbf{\Lambda}_{x})^{-1/4}\mathbf{\Lambda}_{x}'\mathbf{\Sigma}_{x}^{-1}\mathbf{x}$$
$$= (\mathbf{\Lambda}_{x}'\mathbf{\Sigma}_{x}^{-1}\mathbf{\Lambda}_{x})^{-1/2}\mathbf{\Lambda}_{x}'\mathbf{\Sigma}_{x}^{-1}\mathbf{x}, \tag{12}$$

with $E(\mathbf{P}_{C\xi}\mathbf{P}_{C\xi}') = (\mathbf{\Lambda}_{x}'\mathbf{\Sigma}_{x}^{-1}\mathbf{\Lambda}_{x})^{-1/2}\mathbf{\Lambda}_{x}'\mathbf{\Sigma}_{x}^{-1}\mathbf{\Lambda}_{x}(\mathbf{\Lambda}_{x}'\mathbf{\Sigma}_{x}^{-1}\mathbf{\Lambda}_{x})^{-1/2} = \mathbf{I}.$ This is the orthogonal factor score proposed by Takeuchi, Yanai, and Mukherjee (1982). As this score is already standardized, the correlation-preserving score for $\mathbf{\Phi} \neq \mathbf{I}$ can simply be computed by a pre-multiplication of Takeuchi et al.'s factor score with $\mathbf{\Phi}^{1/2}$, so that

$$\mathbf{P}_{C\xi}^{*} = \mathbf{\Phi}^{1/2}(\mathbf{\Lambda}_{x}'\mathbf{\Sigma}_{x}^{-1}\mathbf{\Lambda}_{x})^{-1/2}\mathbf{\Lambda}_{x}'\mathbf{\Sigma}_{x}^{-1}\mathbf{x}. \tag{13}$$

Note that Equations 11 and 13 describe the relationship between mean plausible values and correlation-preserving scores for $\mathbf{P}_{\xi} = \mathbf{F}_{\xi}^{R}$, which depends on the number of imputations. For a small number of imputations $\mathbf{P}_{\xi} \neq \mathbf{F}_{\xi}^{R}$ so that $\mathbf{\Phi}^{-1/2}\mathbf{P}_{C\xi}^{*} \neq \mathbf{P}_{C\xi}$.

However, for any factor score and for any mean plausible value the determinacy should be computed. According to Beauducel and Hilger (in press, Eq. 7) the determinacy of the mean plausible values $\mathbf{P}_{C\xi}$ for exogenous factors can be estimated by means of

$$\mathbf{D}_{C\xi} = diag(\mathbf{P}_{C\xi}\mathbf{P}_{C\xi}')^{-1/2}\mathbf{P}_{C\xi}\mathbf{x}'\mathbf{\Sigma}_{x}^{-1}\mathbf{\Lambda}_{x}\mathbf{\Phi}, \tag{14}$$

and the determinacy of mean plausible values for endogenous factors can be estimated by means of

$$\mathbf{D_{C\eta}} = diag\,(\mathbf{P_{C\eta}}\mathbf{P_{C\eta}^{'}})^{-1/2}\mathbf{P_{C\eta}}\mathbf{y}^{'}\mathbf{\Sigma_y^{-1}}\mathbf{\Lambda_y}(\mathbf{\Gamma\Phi\Gamma^{'}} + \mathbf{\Psi}). \tag{15}$$

## Example

A simulated data set containing $n$ = 10,000 cases, 15 normally distributed N(0,1) observed variables ($\mathbf{x}$) as a measurement model of three exogenous factors $\mathbf{\xi}$ and 10 normally distributed N(0,1) observed variables ($\mathbf{y}$) as a measurement model for two endogenous factors $\mathbf{\eta}$ were generated with IBM SPSS Version 26. The data file (csv) can be found in the supplement. BSEM was performed with Mplus 8.4 (Muthén & Muthén, 2019) in order to estimate the model parameters of the conceptual model presented in Figure 1 (Mplus syntax-file in Supplements).

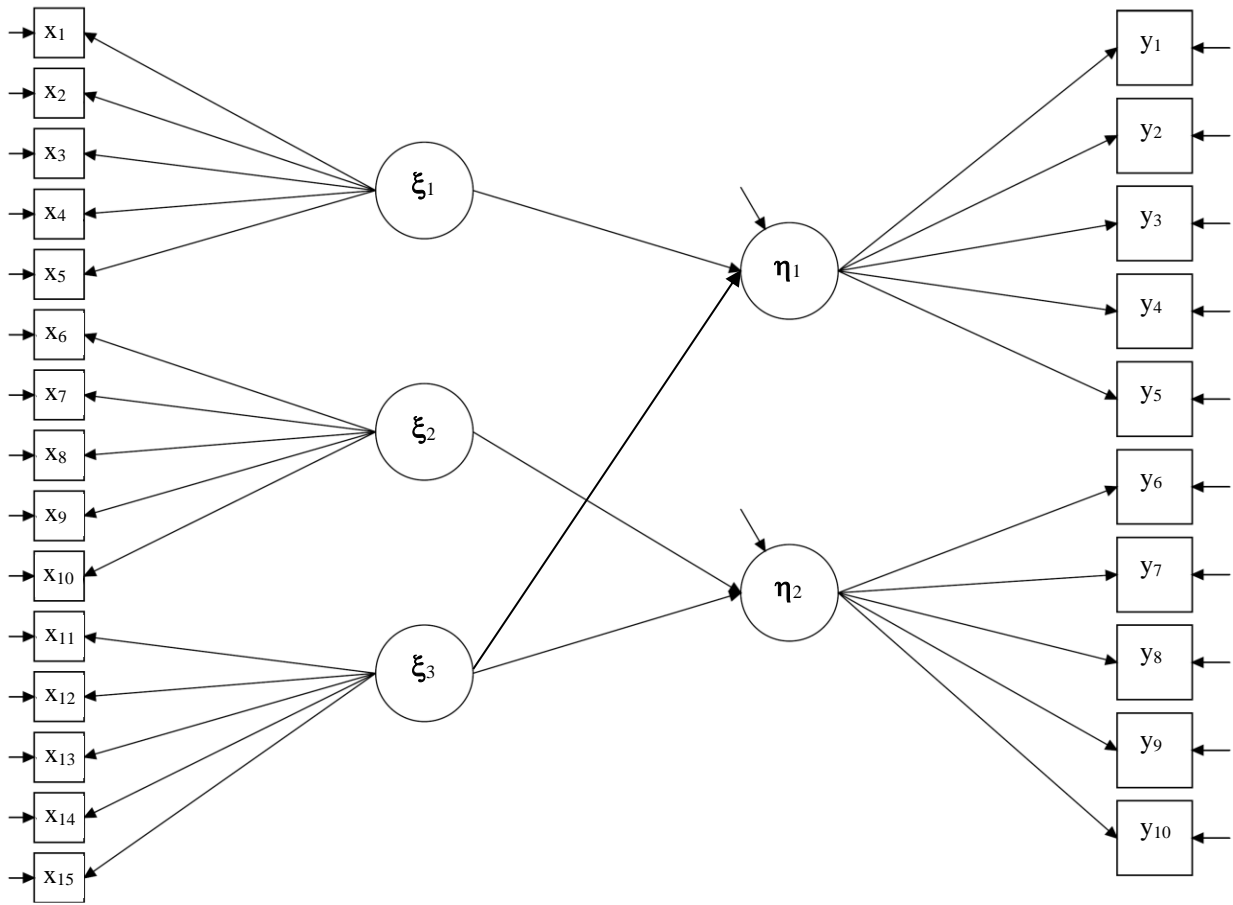

Figure 1. Conceptual model with three exogenous and two endogenous factors. The Bayesian model parameters are given in Table 1.

Factor variances were fixed to one and for each factor five salient loadings were freely estimated and non-salient loadings were estimated with normally distributed priors with a zero mean and a variance of $\sigma^2 = 0.01$. As no model-misfit was simulated, the model fit was excellent (95th confidence-interval for difference between observed and replicated $\chi^2$ [1670.88, 1798.32], posterior predictive p-value < 0.001, prior posterior predictive p-value < .001, RMSEA = 0.027, CFI = 0.991). The model parameter estimates are given in Table 1.

Table 1. BSEM Model parameter estimates (completely standardized solution)

| x | $\mathbf{\Lambda_x}$ | | | y | $\mathbf{\Lambda_y}$ | |
|---|---|---|---|---|---|---|
| $x_1$ | **0.750** | 0.066 | 0.025 | $y_1$ | 0.160 | 0.251 |
| $x_2$ | **0.845** | 0.049 | 0.002 | $y_2$ | 0.160 | 0.251 |
| $x_3$ | **0.938** | 0.031 | -0.021 | $y_3$ | **0.999** | -0.041 |
| $x_4$ | **0.845** | 0.049 | 0.002 | $y_4$ | **0.999** | -0.041 |
| $x_5$ | **0.845** | 0.049 | 0.002 | $y_5$ | **0.999** | -0.041 |
| $x_6$ | 0.031 | **0.762** | 0.023 | $y_6$ | -0.038 | **0.534** |
| $x_7$ | 0.008 | **0.858** | 0.001 | $y_7$ | -0.038 | **0.534** |
| $x_8$ | -0.015 | **0.953** | -0.022 | $y_8$ | -0.037 | **0.533** |
| $x_9$ | 0.008 | **0.859** | 0.000 | $y_9$ | -0.038 | **0.533** |
| $x_{10}$ | 0.008 | **0.858** | 0.001 | $y_{10}$ | -0.038 | **0.534** |
| $x_{11}$ | 0.064 | 0.027 | **0.749** | | $E(\mathbf{\eta\eta'})$ | |
| $x_{12}$ | 0.046 | 0.002 | **0.846** | | 1.000 | **0.513** |
| $x_{13}$ | 0.029 | -0.024 | **0.942** | | **0.513** | 1.000 |
| $x_{14}$ | 0.047 | 0.002 | **0.846** | | | |
| $x_{15}$ | 0.047 | 0.002 | **0.846** | | $\mathbf{\Gamma}$ | |
| | $\mathbf{\Phi}$ | | | | $\eta_1$ | $\eta_1$ |
| | 1.000 | 0.275 | 0.270 | $\xi_1$ | 0.270 | 0.000 |
| | 0.275 | 1.000 | 0.324 | $\xi_2$ | 0.000 | 0.037 |
| | 0.270 | 0.324 | 1.000 | $\xi_3$ | 0.016 | **0.447** |

*Note.* Model parameters greater .40 are given in bold face.

The path coefficients for the prediction of the mean plausible values of the endogenous factors by the mean plausible values of the latent exogenous factors are given in Table 2, together with the path coefficients for the correlation-preserving mean plausible values. The SPSS Syntax for the computation of the mean plausible values and the corresponding regression analyses are given in the Appendix.

Table 2. Standardized path coefficients (beta) based on mean plausible values and on the basis of correlation-preserving mean plausible values.

|  | $P_{\eta 1}$ | $P_{\eta 2}$ |  | $P_{C\eta 1}$ | $P_{C\eta 2}$ |
|---|---|---|---|---|---|
| $P_{\xi 1}$ | 0.275 | -0.038 | $P_{C\xi 1}$ | 0.270 | 0.000 |
| $P_{\xi 2}$ | -0.079 | 0.005 | $P_{C\xi 2}$ | 0.000 | 0.037 |
| $P_{\xi 3}$ | 0.053 | 0.549 | $P_{C\xi 3}$ | 0.016 | 0.447 |

Only for the correlation-preserving mean plausible values the path coefficients are identical to the path coefficients of the model (Table 1). The coefficients of determinacy are given in Table 3. They are very similar for the mean plausible values and for the correlation-preserving mean plausible values.

Table 3. Coefficients of determinacy for mean plausible values and for correlation-preserving mean plausible values.

| $P_{\xi 1}$ | $P_{\xi 2}$ | $P_{\xi 3}$ | $P_{\eta 1}$ | $P_{\eta 2}$ |
|---|---|---|---|---|
| .97 | .97 | .97 | .97 | .85 |
| $P_{C\xi 1}$ | $P_{C\xi 2}$ | $P_{C\xi 3}$ | $P_{C\eta 1}$ | $P_{C\eta 2}$ |
| .97 | .97 | .97 | .99 | .82 |

## Discussion

As mean plausible values based on 500 or more imputations are a proxy of the regression factor score, which does not preserve the inter-correlations of the factors (Skrondal & Laake, 2001), it was concluded that mean plausible values are not correlation-preserving. It might, however, be of interest in the context of BSEM to compute correlation-preserving mean plausible values with the same inter-correlations as the factors. Only correlation-preserving mean-plausible values will result in the same path coefficients from exogenous factors to endogenous factors as the model. Therefore, correlation-preserving mean plausible values were proposed. An example demonstrates how correlation-preserving mean plausible values can be computed from mean plausible values. It is also shown that only the correlation-preserving mean plausible values yield the same path coefficients as the factors.

Factor score determinacies of mean plausible values and of correlation-preserving mean plausible values were similar. It should be noted that the determinacy of correlation-preserving mean plausible values will typically be smaller than the determinacy of the mean plausible values. This follows from the mean plausible value being a proxy of the regression factor score, which has the largest possible determinacy in a given data set. However, as one can see in the example, the prediction of the endogenous factors is different for the mean plausible values and for the correlation-preserving mean plausible values. This may result in a larger determinacy of the correlation-preserving mean plausible values for the endogenous factors. Further research may explore the conditions for the higher determinacy of correlation-preserving mean plausible values for endogenous factors systematically.

When mean plausible values are a proxy of the regression factor score, the correlation-preserving mean plausible values are a proxy of a correlation-preserving version of Takeuchi et al.'s (1982) factor score. As Takeuchi et al.'s factor score has been shown to be identical to Anderson-Rubin's factor score (Anderson & Rubin, 1956; Beauducel, 2015), this also implies that McDonald's (1981) correlation-preserving factor score, will also be equivalent to the correlation-preserving mean plausible values under this condition. Then, mean plausible values need not to be computed and the scores can directly be computed from the model parameters and the observed variables. A limitation of the present study is the focus on the mean plausible values, whereas the median of the plausible values might also be of interest. Whether a correlation-preserving version of the median plausible value might be of interest, especially with small samples might be explored in further studies.

In order to compute the correlation-preserving mean plausible values, the inter-factor correlations estimated by means of BSEM and the mean plausible values should be entered into the example syntax provided in the Appendix. Therefore, the procedure proposed here can also be applied to mean plausible values that are based on a small number of imputations. When the determinacies are to be computed, the loadings and inter-correlations of the exogenous factors, the loadings and inter-correlations of endogenous factors should also be inserted into the syntax.

**Acknowledgement**

This study was funded by the German Research Foundation (DFG) to AB, BE 2443/18-1.

# Appendix

IBM SPSS-Syntax.

```
* Encoding: UTF-8.

* Syntax for the example presented in the manuscript.
* For use in other contexts, enter location and name of the data-file containing mean
  plausible values and adapt the variable number according to your model.

DATA LIST FILE="C:\Example_data_plausible_values.dat" fixed records=1
/1 x1 to x15 (15F6.3)   y1 to y10 (10F6.3)
Ksi1_meanPlausible (F6.3)  Ksi1_medianPlausible (F6.3) Ksi1_SD (F6.3) Ksi1_perc2p5 (F6.3)
Ksi1_perc97p5 (F6.3)
Ksi2_meanPlausible (F6.3)  Ksi2_medianPlausible (F6.3) Ksi2_SD (F6.3) Ksi2_perc2p5 (F6.3)
Ksi2_perc97p5 (F6.3)
Ksi3_meanPlausible (F6.3)  Ksi3_medianPlausible (F6.3) Ksi3_SD (F6.3) Ksi3_perc2p5 (F6.3)
Ksi3_perc97p5 (F6.3)
Eta1_meanPlausible (F6.3)  Eta1_medianPlausible (F6.3) Eta1_SD (F6.3) Eta1_perc2p5 (F6.3)
Eta1_perc97p5 (F6.3)
Eta2_meanPlausible (F6.3)  Eta2_medianPlausible (F6.3) Eta2_SD (F6.3) Eta2_perc2p5 (F6.3)
Eta2_perc97p5 (F6.3).
Dataset name dataset1.
save outfile="C:\Example_data_plausible_values.sav".

MATRIX.

get P_Ksi/variables= Ksi1_meanPlausible Ksi2_meanPlausible  Ksi3_meanPlausible
/file='C:\Example_data_plausible_values.sav'.
get P_Eta/variables= Eta1_meanPlausible Eta2_meanPlausible
/file='C:\Example_data_plausible_values.sav'.

get x/variables=  x1 to x15
/file='C:\Example_data_plausible_values.sav'.
get y/variables=  y1 to y10
/file='C:\Example_data_plausible_values.sav'.

* In the following matrices are the values that are also given in Table 1.
* For use in other contexts, enter the corresponding values from your BSEM-OUTPUT:.

* Loadings of measured variables on Ksi.
compute Lx={
0.750, 0.066, 0.025;
0.845, 0.049, 0.002;
0.938, 0.031,-0.021;
0.845, 0.049, 0.002;
0.845, 0.049, 0.002;
0.031, 0.762, 0.023;
0.008, 0.858, 0.001;
-0.015, 0.953, -0.022;
0.008, 0.859, 0.000;
0.008, 0.858, 0.001;
0.064, 0.027, 0.749;
0.046, 0.002, 0.846;
0.029,-0.024, 0.942;
0.047, 0.002, 0.846;
0.047, 0.002, 0.846}.

* Intercorrelations of Ksi.
compute Phi={
 1.000, 0.275, 0.270;
 0.275, 1.000, 0.324;
 0.270, 0.324, 1.000
}.

* Loadings of measured variables on Eta.
compute Ly={
 0.160, 0.251;
 0.160, 0.251;
 0.999,-0.041;
 0.999,-0.041;
 0.999,-0.041;
```

```
-0.038, 0.534;
-0.038, 0.534;
-0.037, 0.533;
-0.038, 0.533;
-0.038, 0.534
}.

* Path coefficients from Ksi to Eta.
compute Gamma={
0.270,  0.000;
0.000,  0.037;
0.016,  0.447
}.

* Intercorrelations of Eta.
compute Ceta = {
1.000, 0.513;
0.513, 1.000
}.

* Computations.
compute P_Ksi=t(P_Ksi).
compute ncases=ncol(P_ksi).
* Mean-centering of P_Ksi.
compute mP=RSUM(P_Ksi)&/ncases.
compute ones=make(nrow(P_Ksi),ncol(P_Ksi),1).
compute mmP=Mdiag(mP)*ones.
compute P_Ksi=P_Ksi-mmP.

compute P_Eta=t(P_Eta).
* Mean-centering of P_Eta.
compute mP=RSUM(P_Eta)&/ncases.
compute ones=make(nrow(P_Eta),ncol(P_Eta),1).
compute mmP=Mdiag(mP)*ones.
compute P_Eta=P_Eta-mmP.

compute P={P_Ksi;P_Eta}.

compute x=t(x).
compute y=t(y).

compute C_P=INV(Mdiag(diag( P*t(P)&/(ncases-1) ))&**0.5) * P*t(P)&/(ncases-1)
* INV(Mdiag(diag( P*t(P)&/(ncases-1) ))&**0.5).
print C_P/format=F5.2/Title="Correlation of mean plausible values".

CALL Eigen(C_P, vec, eig).
compute C_P12=vec*Mdiag(eig)&**0.5*t(vec).

compute Gamma=t(Gamma).
compute Cetaksi=(Gamma)*Phi.
compute tcetaks=t(Cetaksi).

compute C={
Phi, tCetaks;
Cetaksi, Ceta }.

Print C/format=F5.2/Title="Correlation of factors according to the model parameters of BSEM".

CALL Eigen(C, vec, eig).
compute C12=vec*Mdiag(abs(eig))&**0.5*t(vec).

* Compute correlation-preserving plausible values according to Equation 10.
compute Pc=C12*INV(C_P12)*INV(Mdiag(diag( P*t(P)&/(ncases-1) ))&**0.5) *P.

print {INV(Mdiag(diag( Pc*t(Pc) )&/(ncases-1))&**0.5)*Pc*t(Pc)&/(ncases-1)
*INV(Mdiag(diag( Pc*t(Pc) )&/(ncases-1))&**0.5)}
/format=F5.2/Title="Check: Correlation of correlation-preserving mean plausible values. Should
be equal to correlation of factors".

* Determinacy.

compute Tdelta=Mdiag(diag( 1-Lx*Phi*t(Lx) )).
compute Sig_x=Lx*Phi*t(Lx) + Tdelta.
```

```
compute Tepsi=Mdiag(diag(1 - Ly*Ceta*t(Ly))).
compute Sig_y=Ly*Ceta*t(Ly) + Tepsi.

* Compute Determinacy of mean plausible values for Ksi.
compute D_Ksi=INV(Mdiag(diag( P_Ksi*t(P_Ksi)&/(ncases-1) )))&**0.5 * P_Ksi
* t(x)&/(ncases-1) * INV(Sig_x)*Lx*Phi.
print {t(diag(D_Ksi))} /format=F5.2/Title="Determinacy of mean plausible values for Ksi".

* Compute Determinacy of mean plausible values for Eta.
compute D_Eta=INV(Mdiag(diag( P_Eta*t(P_Eta)&/(ncases-1) ))) * P_Eta
*T(y)&/(ncases-1) * INV(Sig_y)*Ly*Ceta.
print {t(diag(D_Eta))} /format=F5.2/Title="Determinacy of mean plausible values for Eta".

* Compute Determinacy of correlation-preserving mean plausible values for Ksi (according
Equation 14).
compute PcKsi = {Pc(1,:);Pc(2,:);Pc(3,:)}.
compute D_cKsi=INV(Mdiag(diag( PcKsi*t(PcKsi)&/(ncases-1) )))&**0.5 *PcKsi
*T(x)&/(ncases-1)*INV(Sig_x)*Lx*Phi.
print {t(diag(D_cKsi))} /format=F5.2/Title="Determinacy of correlation-preserving mean plausible
values for Ksi (according to Equation 12)".

* Compute Determinacy of correlation-preserving mean plausible values for Eta (according
Equation 15).
compute PcEta = {Pc(4,:);Pc(5,:)}.
compute D_cEta=INV(Mdiag(diag( PcEta*t(PcEta)&/(ncases-1) ))) * PcEta
*T(y)&/(ncases-1)*INV(Sig_y)*Ly*Ceta.
print {t(diag(D_cEta))} /format=F5.2/Title="Determinacy of correlation-preserving mean plausible
values for Eta (according to Equation 13)".

save {t(Pc)}/outfile="C:\Example_data_correlation_preserving_plausible_values.sav"/variables
Pc_Ksi1 Pc_Ksi2 Pc_Ksi3 Pc_Eta1 Pc_Eta2.

END MATRIX.

* The following regression analyses are performed in order to check whether the
  correlation-preserving mean plausible values yield the same standardized
  coefficients as the BSEM model.

* Compare regression-coefficients of conventional mean plausible values...

Dataset activate Dataset1.
REGRESSION  /MISSING LISTWISE  /STATISTICS COEFF OUTS R ANOVA
  /CRITERIA=PIN(.05) POUT(.10)  /NOORIGIN   /DEPENDENT Eta1_meanPlausible
  /METHOD=ENTER Ksi1_meanPlausible Ksi2_meanPlausible  Ksi3_meanPlausible.

Dataset activate Dataset1.
REGRESSION  /MISSING LISTWISE  /STATISTICS COEFF OUTS R ANOVA
  /CRITERIA=PIN(.05) POUT(.10)  /NOORIGIN   /DEPENDENT Eta2_meanPlausible
  /METHOD=ENTER Ksi1_meanPlausible Ksi2_meanPlausible  Ksi3_meanPlausible.

* …with regression-coefficients of correlation-preserving mean plausible values:.

get file="C:\Example_data_correlation_preserving_plausible_values.sav".
Dataset name Dataset2.

Dataset activate Dataset2.
REGRESSION  /MISSING LISTWISE  /STATISTICS COEFF OUTS R ANOVA
  /CRITERIA=PIN(.05) POUT(.10)  /NOORIGIN   /DEPENDENT Pc_Eta1
  /METHOD=ENTER Pc_Ksi1 Pc_Ksi2 Pc_Ksi3.

Dataset activate Dataset2.
REGRESSION  /MISSING LISTWISE  /STATISTICS COEFF OUTS R ANOVA
  /CRITERIA=PIN(.05) POUT(.10)  /NOORIGIN   /DEPENDENT Pc_Eta2
  /METHOD=ENTER Pc_Ksi1 Pc_Ksi2 Pc_Ksi3.
```